\documentclass[aps,pra,amsmath,amsfonts,amssymb,twocolumn,superscriptaddress,showpacs]{revtex4}

\usepackage{amssymb}
\usepackage{graphicx,psfrag,color}
\usepackage{dcolumn}
\usepackage{bm}
\usepackage{mathbbol}

\begin{document}
\flushbottom

\title{Entangling Power of Disordered Quantum Walks}

\author{Rafael Vieira}
\author{Edgard P. M. Amorim}
\email{eamorim@joinville.udesc.br}
\affiliation{Departamento de F\'isica, Universidade do Estado de
Santa Catarina, 89219-710, Joinville, SC, Brazil}
\author{Gustavo Rigolin}
\email{rigolin@ufscar.br}
\affiliation{Departamento de F\'isica, Universidade Federal de
S\~ao Carlos, 13565-905, S\~ao Carlos, SP, Brazil}

\date{\today}

\begin{abstract}
We investigate how the introduction of different types of disorder affects the generation of entanglement between the internal (spin) and
external (position) degrees of freedom in one-dimensional
quantum random walks ($QRW$). Disorder is modeled by adding another random feature to $QRW$, i.e., the quantum coin that drives 
the system's evolution is randomly chosen at each position and/or at each time step, giving rise to either
dynamic, fluctuating, or static disorder. The first one is position-independent, with every lattice site having the same coin at a
given time, the second has time and position dependent randomness, while the third one is time-independent. We show for several
levels of disorder that dynamic disorder is the most powerful entanglement generator, 
followed closely by fluctuating disorder. Static disorder is the less efficient entangler, 
being almost always less efficient than the ordered case. Also, dynamic and fluctuating disorder lead to maximally entangled states
asymptotically in time for any initial condition while static disorder has no asymptotic limit and, similarly to
the ordered case, has a long time behavior highly sensitive to the initial conditions.
\end{abstract}


\maketitle

\section{Introduction} 

The simplest generalization of the classical random walk ($CRW$) \cite{per05ray05} 
to the quantum domain is the one-dimensional discrete time quantum
random walk ($QRW$) \cite{aha93} based on two-level systems (qubits) \cite{kem03,and12}. 
In this model the qubit moves one step to the right or to the left according to its internal degree 
of freedom (spin state). If its spin is up, for example, it moves to the right and
if its spin is down it moves to the left.  At each step of the walk a unitary operation
(``quantum coin'') $C$ is applied to the qubit's internal degree of freedom prior to
the action of the displacement operator $S$, which moves the qubit as described above.

This apparently simple quantum dynamic model has transport properties completely different
from its classical counterpart, the causes of which are mainly due to the principle of 
superposition of quantum mechanics. Indeed, the dynamics of $QRW$ creates a superposition 
among the position states (external degree of freedom) of the system, opening the way to interference
effects to take place that ultimately determine the transport properties of $QRW$. 
For instance, the position probability distribution $P(j)$ for certain types
of $QRW$ is far from being a Gaussian distribution \cite{kem03,and12}, as is the case for $CRW$. 
Moreover, for most $QRW$ we have a ballistic behavior for the displacement of the walker, 
i.e., the variance $\sigma^2$ of $P(j)$ is proportional to $n^2$,
with $n$ the number of steps of the walk. For $CRW$, however, we have a diffusive 
behavior ($\sigma^2 \propto n$).  
Another characteristic of $QRW$ is the generation of entanglement between the internal and external degrees of 
freedom of the qubit during its time evolution \cite{car05,aba06,sal12}.
This is a genuine quantum feature of $QRW$ since there is no analog to entanglement for $CRW$.  

The transport and entanglement properties of $QRW$ were extensively studied for
the ordered case \cite{aha93,kem03,and12,bos09,goy10,shi10,fra11,all12,rol13,mou13,shi13,shi14}, 
where the quantum coin is fixed throughout the time evolution 
or changes in time in a predictable way. Most ordered $QRW$ have ballistic transport and their asymptotic (long time) 
entanglement is highly dependent on the initial condition of the walker and almost never reaches its maximal possible value.
Ordered $QRW$ has also proved useful in a possible implementation of quantum search algorithms \cite{she03}, of a universal quantum 
computer \cite{chi09,lov10}, and in the simulation of Dirac-like Hamiltonians \cite{cha13}.
For disordered $QRW$, where the coin and/or the displacement operator is randomly chosen at each lattice site and/or time step,
only the transport properties were studied in some depth 
\cite{rib04,ban06,bue04,woj04,rom05,sha03,joy11,ahl11a,ahl11b,ahl12,gho14,nic14}. In this scenario,
depending on the type of disorder, we can have diffusive or ballistic transport or Anderson localization. 

\begin{widetext}
\begin{figure*}[!ht]
\includegraphics[width=18cm]{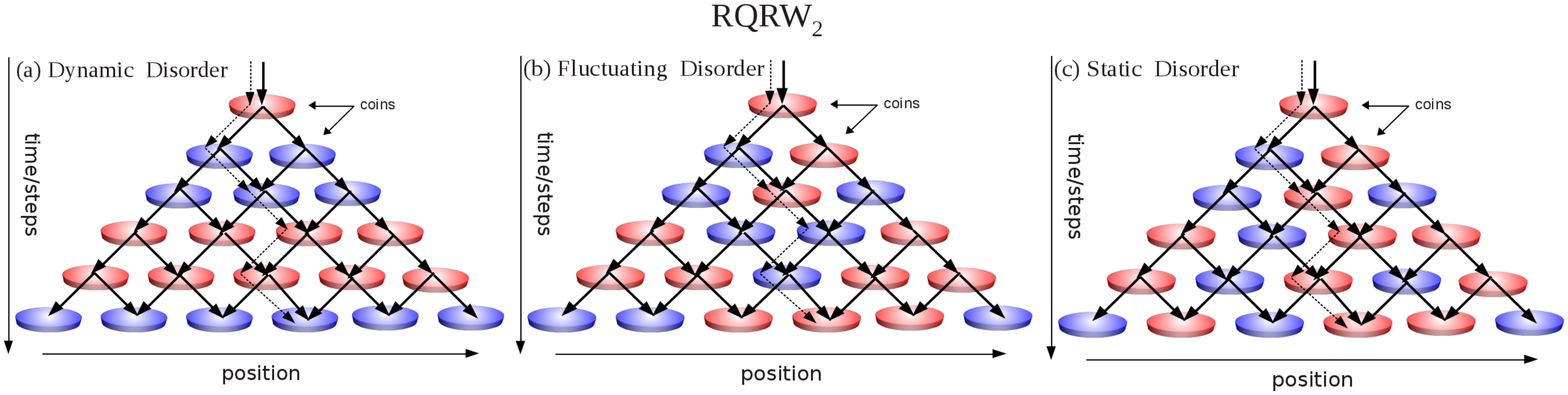}
\includegraphics[width=18cm]{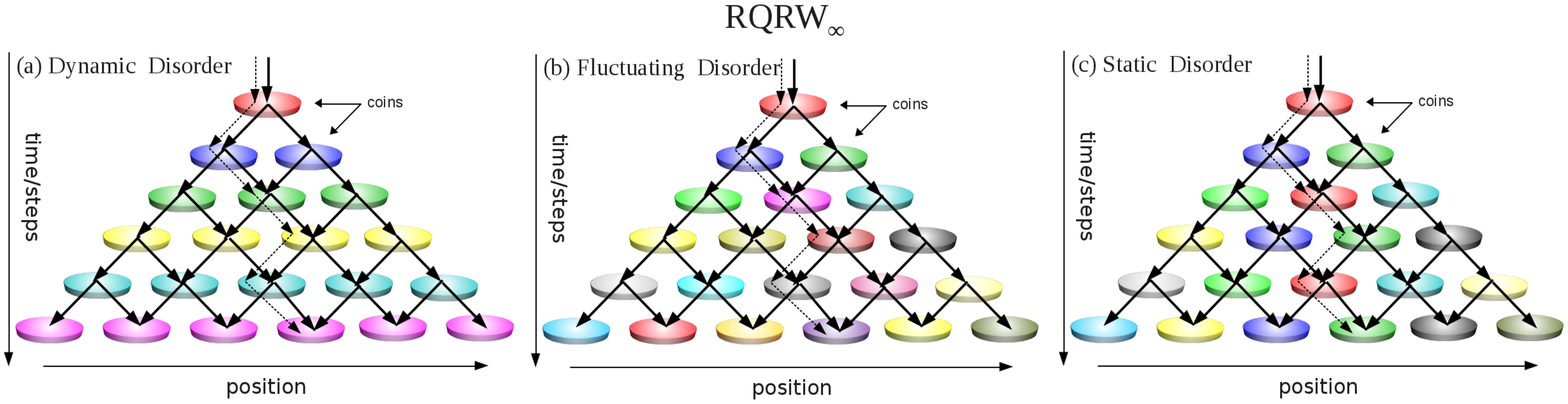}
\caption{\label{fig0}(color online) The dashed lines
represent one of many possible realizations of $CRW$ 
and the solid curves the probability amplitudes associated to $RQRW.$
Upper panel: $RQRW_2$ with only two coins $C(j,t)$ (red and
blue discs). Lower panel: $RQRW_\infty$ with $C(j,t)$
chosen randomly from continuous uniform distributions of
quantum coins. In all panels for dynamic disorder (left) we have 
$C(j,t)=C(t)$, for fluctuating disorder (middle) $C(j,t)$, and for 
the static case $C(j,t)=C(j)$. Pictorially, for dynamic disorder 
all coins are the same within a given row (same coin everywhere at time $t$)
while for static disorder they are equal within a given column (same coin at a 
given position $j$ for all time). For fluctuating disorder, at each time step 
the coin changes randomly at each position and independently of the other sites.}
\end{figure*}
\end{widetext}

The first systematic study of the entanglement properties of disordered $QRW$ was carried out in Ref. \cite{vie13}, where dynamic
disorder was employed. A dynamically disordered $QRW$ is such that the coin $C$ is the same at all lattice sites but 
changes randomly at each time step (see Fig. \ref{fig0}). The main results of \cite{vie13} and its supplemental material 
were analytical and numerical proofs that (i) 
the asymptotic entanglement between the position and the spin degrees of freedom are maximal and (ii) independent 
of the initial state of the walker. These facts were surprising for the introduction of noise (disorder) to $QRW$ enhanced
its entangling power considerably for any initial condition, against naive expectations that noise would wash out
the quantum features of $QRW$.

In addition to dynamic disorder, there exist static and fluctuating disorder 
\cite{sch10,sch11,cre13}. In the static case the coins are fixed in time but randomly distributed along the lattice sites.
Fluctuating disorder combines both features of the dynamic and static case: the coins change randomly at each time step and at each
lattice site (see Fig. \ref{fig0}). So far, the only work studying the entanglement behavior of statically and dynamically disordered
$QRW$ is Ref. \cite{cha12}, where it is shown for only one initial condition and for a $100$-step walk that dynamic disorder outperforms
both the ordered and the static case, with the latter giving the worst result. 

What would happen, though, for other initial conditions? What would one get by employing fluctuating disorder?
What sort of disorder is the most powerful entanglement generator? How all these different types of disorder compare to the ordered case?
Our main goal in this article is two answer the previous questions by systematically investigating the entangling power 
of $QRW$ subjected to dynamic, fluctuating, and static disorder. Specifically, we want to test if all types of disorder lead to an
asymptotic limit and how it is approached (in case there is one). Also, we want to determine
the asymptotic behavior of the entanglement for these three types of disorder and also its dependence 
(or independence) on the initial conditions of the walker.

\section{Disordered Quantum Walks}

Let us start by presenting the notation and nomenclature employed in the rest of this
article. Following Ref. \cite{vie13} we call a disordered $QRW$ a \textit{random 
quantum random walk}, or $RQRW$ for short. The extra $R$ simply reminds us that
there is a new random aspect in the disordered case when compared to the ordered one,
namely, the quantum coin changes in an unpredictable way. Also, we will be
dealing with walks where we either have only two or
a continuum of quantum coins to play with. In
order to keep track of these different situations, we call the former 
process $RQRW_2$ and the latter $RQRW_\infty$, with the subscripts
denoting that we have, respectively, two or an infinity of coins at our
disposal.

The Hilbert space describing a $RQRW$ can be written as $\mathcal{H}=\mathcal{H}_C\otimes \mathcal{H}_P$.
Here $\mathcal{H}_C$ is a complex vector space of dimension two associated to the internal degree
of freedom of the qubit, namely, its spin state, that points either up, $|\!\uparrow\rangle$, or down, $|\!\downarrow\rangle$;  
$\mathcal{H}_P$ represents a countable infinite-dimensional complex Hilbert space, whose 
base is given by the kets $|j\rangle$, $j\in
\mathbb{Z}$ (integers). Each ket is associated to the position $j$ of the qubit on a
one-dimensional lattice. 

A general initial state for the walker can be written as
\begin{equation}
|\Psi(0)\rangle =
\sum_{j=-\infty}^{\infty}[a(j,0)|\uparrow\rangle \otimes|j\rangle+b(j,0)|\!\downarrow\rangle\otimes|j\rangle],
\end{equation}
with normalization condition demanding that $\sum_j(|a(j,0)|^2+|b(j,0)|^2)=1$. The time $t$ is discrete and
changes in increments of one. In an $n$-step walk the time goes from $t=0$ to $t=n$. 
For the most general $RQRW$, after $n$ steps the state describing the qubit is 
\begin{equation}
|\Psi(n)\rangle = U(n) \cdots
U(1)|\Psi(0)\rangle=\mathcal{T}\prod_{t=1}^{n}U(t)|\Psi(0)\rangle,
\end{equation}
where $\mathcal{T}$ represents a time-ordered product, and
\begin{eqnarray}
U(t) &=& S\mathcal{C}(t), \label{evolution} \\
S&=&\hspace{-.35cm}\sum_{j=-\infty}^{\infty}\hspace{-.2cm}(|\!\uparrow\rangle\langle\uparrow\!|\otimes\!|j+1\rangle\langle j|
+|\!\downarrow\rangle\langle\downarrow\!|\otimes\!|j-1\rangle\langle j|), \label{displacement}\\
\mathcal{C}(t)&=&\sum_{j=-\infty}^{\infty}C(j,t)\otimes |j\rangle\langle j|.
\label{coin}
\end{eqnarray}
Here $C(j,t)$ is the quantum coin acting on the internal degree of freedom at site $j$ and at time $t$
and $S$ is the conditional displacement operator. As outlined in the introductory
section, $S$ moves a qubit located in $j$ to $j+1$ if its spin is up and
to $j-1$ if its spin is down.

An arbitrary $SU(2)$ unitary operator gives the most general 
$C(j,t)$ and up to an irrelevant global phase it can be written as 
\begin{eqnarray}
C(j,t)&=&c_{\uparrow\uparrow}(j,t)|\uparrow\rangle\langle\uparrow| +
c_{\uparrow\downarrow}(j,t) |\uparrow\rangle\langle\downarrow| \nonumber \\
&&+c_{\downarrow\uparrow}(j,t)|\downarrow\rangle\langle\uparrow| +
c_{\downarrow\downarrow}(j,t) |\downarrow\rangle\langle\downarrow| \nonumber \\
&=&
\left(
\begin{array}{cc}
c_{\uparrow\uparrow}(j,t) & c_{\uparrow\downarrow}(j,t) \\
c_{\downarrow\uparrow}(j,t) & c_{\downarrow\downarrow}(j,t)
\end{array}
\right)
\nonumber \\
&=&
\hspace{-.2cm}\left(\hspace{-.2cm}
\begin{array}{cc}
\sqrt{q(j,t)} & \sqrt{1-q(j,t)}e^{i\theta(j,t)} \\
\sqrt{1-q(j,t)}e^{i\varphi(j,t)} & -\sqrt{q(j,t)}e^{i[\theta(j,t)+\varphi(j,t)]}
\end{array}
\hspace{-.2cm}\right),\hspace{-.2cm} \nonumber \\
\label{qcoin}
&&
\end{eqnarray}
with  $0\leq q(j,t)\leq1$ and $0\leq\theta(j,t),\varphi(j,t)\leq 2\pi$.

The parameter $q(j,t)$ dictates the bias of
the quantum coin. If $q(j,t)=1/2$ the coin generates an equal superposition of
the spin states when acting on the states $|\!\uparrow\rangle$ or
$|\!\downarrow\rangle$. If $q(t)\neq 1/2$ we have an uneven superposition of
spin up and down. The other two parameters, $\theta(j,t)$ and $\varphi(j,t)$, 
control the relative phase between the two
states in the superposition.

For $RQRW_2$ (see Fig. \ref{fig0}) only two quantum coins are available and hence
$C(j,t)= C_1$ or $C_2$. At each step and/or position of the walk the choice between $C_1$
or $C_2$ is made according to the result of a classical coin tossing. 
For $RQRW_\infty$ we deal with an infinite number of
$C(j,t)$ and at each step and/or position $C(j,t)$ is defined as follows. 
The three independent parameters $q(j,t)$, $\theta(j,t)$, and $\varphi(j,t)$ of $C(j,t)$ 
are chosen at each step and/or position from three distinct 
uniform continuous distributions defined over their allowed values.  

The time evolution of the system is obtained applying $U(t)$, Eq.~(\ref{evolution}),
to an arbitrary state at time $t-1$,
\begin{eqnarray}
|\Psi(t)\rangle &=& U(t)|\Psi(t-1)\rangle \nonumber \\
&=&\sum_{j=-\infty}^{\infty}[a(j,t)|\!\uparrow\rangle|j\rangle+b(j,t)|\!\downarrow\rangle|j\rangle], 
\end{eqnarray}
where
\begin{eqnarray}
a(j,t) &=& c_{\uparrow\uparrow}(j-1,t)a(j-1,t-1) \nonumber \\
&&+ c_{\uparrow\downarrow}(j-1,t)b(j-1,t-1), \nonumber \\
b(j,t) &=& c_{\downarrow\uparrow}(j+1,t)a(j+1,t-1) \nonumber \\
&&+ c_{\downarrow\downarrow}(j+1,t)b(j+1,t-1).
\label{recurrence}
\end{eqnarray}

These are the recurrence relations we employ to numerically simulate the results in the following sections and before we move on
it is important to emphasize the key differences in the equations of motion above for the three types of 
disorder.

\subsection{Dynamic Disorder}

In this case $C(j,t)=C(t)$, i.e., the quantum coin is the same for all position $j$ and changes randomly from one time step to the other
(see left panels of Fig.~\ref{fig0}).
Therefore, Eq.~(\ref{coin}) can be written as $\mathcal{C}(t)=\sum_{j=-\infty}^{\infty}C(j,t)\otimes |j\rangle\langle j|
=C(t)\otimes\sum_{j=-\infty}^{\infty} |j\rangle\langle j|=C(t)\otimes\mathbb{1}_P$, where $\mathbb{1}_P$ is the identity operator
defined over $\mathcal{H}_P$. In this case Eq.~(\ref{evolution}) is simply $U(t)=S[C(t)\otimes\mathbb{1}_P]$ and the recurrence relations
become those given in \cite{vie13}
\begin{eqnarray}
a(j,t) \hspace{-.1cm} &=& \hspace{-.1cm} c_{\uparrow\uparrow}(t)a(j-1,t-1) + c_{\uparrow\downarrow}(t)b(j-1,t-1), \nonumber \\
b(j,t) \hspace{-.1cm} &=& \hspace{-.1cm} c_{\downarrow\uparrow}(t)a(j+1,t-1) + c_{\downarrow\downarrow}(t)b(j+1,t-1).
\label{recurrenced}
\end{eqnarray}

\subsection{Fluctuating Disorder}

Here we have a position-dependent quantum coin that is not fixed throughout the time evolution, i.e., at each time step new coins are drawn
at each position $j$ (see middle panels of Fig.~\ref{fig0}). The time evolution is given by Eqs.~(\ref{evolution})-({\ref{coin}}) and the
recurrence relations are those given in Eq.~(\ref{recurrence}).

\subsection{Static Disorder}

In this scenario the coins are fixed during the time evolution, $C(j,t)=C(j)$, but randomly drawn at each site $j$
(see right panels of Fig.~\ref{fig0}). The recurrence relations
become
\begin{eqnarray}
a(j,t) \hspace{-.1cm} &=& \hspace{-.1cm} c_{\uparrow\uparrow}(j\hspace{-.05cm}-\hspace{-.05cm}1)a(j\hspace{-.05cm}-\hspace{-.05cm}1,t-1) + 
c_{\uparrow\downarrow}(j\hspace{-.05cm}-\hspace{-.05cm}1)b(j\hspace{-.05cm}-\hspace{-.05cm}1,t-1), \nonumber \\
b(j,t) \hspace{-.1cm} &=& \hspace{-.1cm} c_{\downarrow\uparrow}(j\hspace{-.05cm}+\hspace{-.05cm}1)a(j\hspace{-.05cm}+\hspace{-.05cm}1,t-1) + 
c_{\downarrow\downarrow}(j\hspace{-.05cm}+\hspace{-.05cm}1)b(j\hspace{-.05cm}+\hspace{-.05cm}1,t-1).\nonumber \\
\label{recurrences}
\end{eqnarray}

For completeness, we just mention that the time-independent 
ordered case is such that the quantum coin is a constant, $C(j,t)=C$, i.e., same coin everywhere
during the whole time evolution.

\section{Entanglement Quantifier}

The total initial state (spin plus position) $|\Psi(0)\rangle$ is always pure in this paper. This, together
with the fact that the evolution of the whole system is unitary, lead to total pure states at any time.  In this scenario,
the natural choice to quantify the entanglement between the internal and external degrees of freedom is the von Neumann 
entropy ($S_E$) of the partially reduced state of either the spin/coin state or the position state, with either choice
furnishing the same entanglement \cite{ben96}. The simplest choice is to work with the spin reduced state $\rho_C(t)=Tr_P[\rho(t)]$, where 
$\rho(t)=|\Psi(t)\rangle\langle\Psi(t)|$ and $Tr_P[\cdot]$ is the trace over the position degree of freedom.

If we write 
$$
\rho_C(t) = 
\alpha(t) |\hspace{-.1cm}\uparrow\rangle\langle\uparrow\hspace{-.1cm}|+
\beta(t)|\hspace{-.1cm}\downarrow\rangle\langle\downarrow\hspace{-.1cm}|+
\gamma(t) |\hspace{-.1cm}\uparrow\rangle\langle\downarrow\hspace{-.1cm}|+ 
\gamma^*(t)|\hspace{-.1cm}\downarrow\rangle\langle\uparrow\hspace{-.1cm}|,
$$
with
$
\alpha(t)=\sum_{j} |a(j,t)|^2, 
\beta(t)=\sum_{j} |b(j,t)|^2, 
\gamma(t)=\sum_{j} a(j,t)b^*(j,t),
$
and $z^*$ the complex conjugate of $z$, 
the entanglement is
\begin{eqnarray}
S_E[\rho(t)]&=&-Tr[\rho_C(t)\log_2\rho_C(t)] \nonumber \\
&=&-\lambda_+(t)\log_2\lambda_{+}(t)-\lambda_{-}(t)\log_2\lambda_{-}(t),
\end{eqnarray}
where $\lambda_{\pm}$ are the eigenvalues of $\rho_C(t)$,
$$
\lambda_{\pm}=\frac{1}{2}{\pm}\sqrt{\frac{1}{4}-\alpha(t)[1-\alpha(t)]+|\gamma(t)|^2}.
$$

Note that we are using the logarithm at base two since we want the maximally entangled state to have $S_E=1$. 
Separable states (no entanglement) have $S_E=0$.

\section{Results}

\subsection{General trends}

Let us start comparing the entangling power among the several types of disorder within the same class
of $RQRW$. The first class is $RQRW_\infty$, where $C$ (Eq.~(\ref{qcoin})) is chosen from an infinity set of possibilities. 
In this case the variables defining $C$, $q \in [0,1]$ and $\theta, \varphi \in [0,2\pi]$, 
are randomly chosen from three distinct uniform distributions defined over their allowed values. 
The second class is $RQRW_2$, where we have two coins only,
the Hadamard ($H$) coin ($q=1/2$, 
$\theta=\varphi=0$) and the Fourier/Kempe ($F$) coin ($q=1/2$,
$\theta=\varphi=\pi/2$). We call $RQRW_2$ balanced if the odds of getting $C_1$ or
$C_2$ at each draw are equal and unbalanced otherwise.

\begin{figure}[!ht]
\includegraphics[width=8cm]{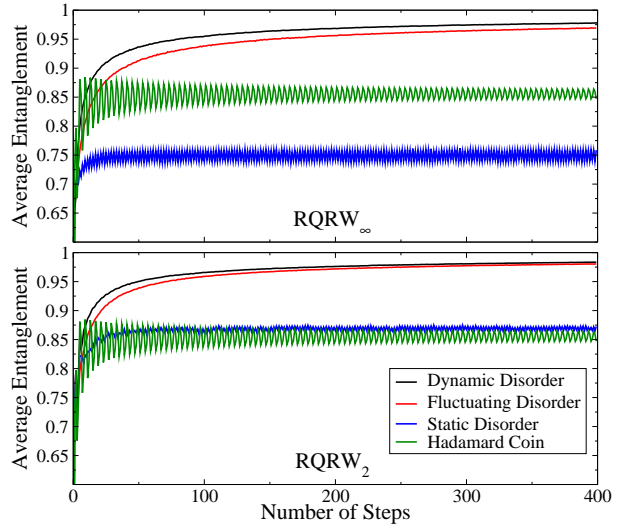}
\caption{\label{fig1}(color online) Average entanglement 
$\langle S_E \rangle$ obtained by averaging over 16,384 initial conditions of the type
$|\Psi(0)\rangle\hspace{-.1cm}=\hspace{-.1cm}(\cos\alpha_s|\hspace{-.1cm}\uparrow\rangle+e^{i\beta_s}
\hspace{-.1cm}\sin\alpha_s|\hspace{-.1cm}\downarrow\rangle) \otimes
(\cos\alpha_p|-1\rangle+e^{i\beta_p}\hspace{-.1cm}\sin\alpha_p|1\rangle)$,
where $\alpha_{s,p}\in [0,\pi]$ and $\beta_{s,p}\in [0,2\pi]$. The
first numerical experiment was implemented with the initial condition
$(\alpha_{s},\beta_{s},\alpha_{p},\beta_p)=(0,0,0,0)$ and the
subsequent ones with quadruples of points in independent increments of $0.4$ until
$\alpha_{s,p} = \pi$ and $\beta_{s,p} = 2\pi$. Here we have a $400$-step
walk. Upper panel: After many steps and from top to bottom we have $RQRW_\infty$ built with dynamic disorder (black),
fluctuating disorder (red), the ordered $QRW$ with the Hadamard coin (green),
and static disorder (blue).
Lower panel: After many steps and from top to bottom we have $RQRW_2$ built with dynamic disorder (black),
fluctuating disorder (red), static disorder (blue), 
and the ordered $QRW$ with the Hadamard coin (green).}
\end{figure}

In Fig. \ref{fig1} we show our first set of numerical experiments, where we have worked with
$i_{max}=16,384$ different initial conditions for both $RQRW_\infty$ (upper panel) and $RQRW_2$ (lower panel)
in a $400$-step walk.
The initial conditions have no entanglement and are delocalized (superposition between positions $|-1\rangle$ and
$|1\rangle$).
For each initial condition $i$ we have implemented the appropriate $RQRW$ and computed at step $t$ 
the entanglement $S^i_E(t)$, where the superscript labels the corresponding initial condition. 
In Fig. \ref{fig1} we plot the average entanglement as a function of the number of steps $t$,
$$
\langle S_E(t) \rangle=\frac{\sum_{i=1}^{i_{max}}S^i_E(t)}{i_{max}},
$$
where the average is taken over all realizations/initial conditions.

Looking at Fig. \ref{fig1} we readily see that for both 
$RQRW_\infty$ and $RQRW_2$ dynamic disorder is the most powerful entangler, followed closely 
by fluctuating disorder. Static disorder, however, is by far the weakest entangler for 
$RQRW_\infty$, even worse than the ordered case. For $RQRW_2$ static disorder is slightly superior than
the ordered case. Moreover, for both classes of $RQRW$	 we can see that 
$\lim_{t\rightarrow \infty}\langle S_E(t)\rangle = 1$  
for the dynamic and fluctuating disorder. For static disorder and order, though, we do not have such
a behavior, with $\langle S_E(t)\rangle$ oscillating about a lower value ($\approx 0.7$ or $\approx 0.8$). 

\begin{figure}[!ht]
\includegraphics[width=8cm]{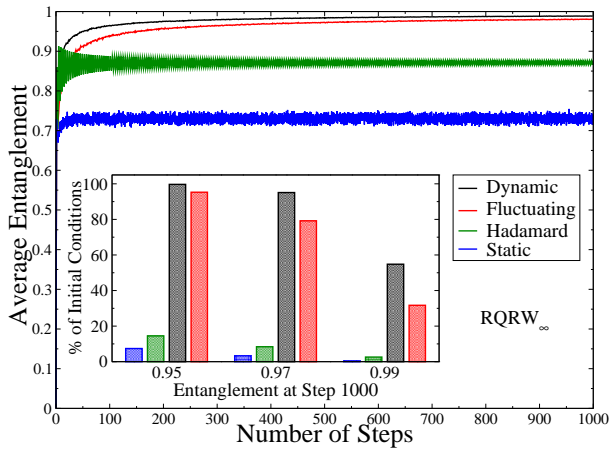}
\caption{\label{fig2}(color online) $\langle S_E \rangle$
was computed averaging over
2,016 localized initial conditions given as
$|\Psi(0)\rangle=(\cos\alpha_s|\hspace{-.1cm}\uparrow\rangle+e^{i\beta_s}
\hspace{-.1cm}\sin\alpha_s|\hspace{-.1cm}\downarrow\rangle)$ $\otimes
|0\rangle$,
with $\alpha_{s}\in [0,\pi]$ and $\beta_{s}\in [0,2\pi]$. The first realization employed the initial
condition $(\alpha_{s},\beta_{s})=(0,0)$ and the following ones
all pairs of points in independent increments of $0.1$ until 
$\alpha_s=\pi$ and $\beta_s=2\pi$. Here we work with $RQRW_\infty$ and a $1000$-step walk.
After many steps and from top to bottom we have dynamic disorder (black), 
fluctuating disorder (red),  the ordered $QRW$ with the Hadamard coin (green),
and static disorder (blue).
The inset gives the rate of initial conditions leading to $S_E$ greater than
$0.95$, $0.97$, and $0.99$ at step $1000$. Note that dynamic disorder outperforms all other types
of disorder in its entanglement generation capability (black bars). Fluctuating disorder (right/red bars)
ranks second while static disorder (left/blue bars)
is the worst entangler, even worse than the ordered case (green bars).}
\end{figure}
\begin{figure}[!ht]
\includegraphics[width=8cm]{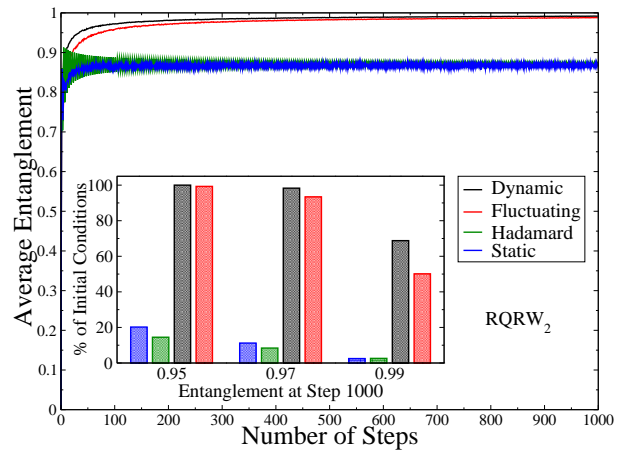}
\caption{\label{fig3}(color online) The same as Fig. \ref{fig2} but now we work
with the balanced $RQRW_2$.
After many steps and from top to bottom we have dynamic disorder (black), 
fluctuating disorder (red), static disorder (blue), and the ordered $QRW$ with the Hadamard coin (green).
Note that the last two curves overlap.
The inset gives the rate of initial conditions leading to $S_E$ greater than
$0.95$, $0.97$, and $0.99$ at step $1000$. Again dynamic disorder outperforms all other types
of disorder in its entanglement generation capability (black bars). Fluctuating disorder (right/red bars)
ranks second while static disorder (left/blue bars) ranks third, this time
being slightly better than the ordered case (green bars).}
\end{figure}

To improve our understanding of the asymptotic limit we now simulate another set of walks with 
$1000$ steps. We work this time with $2,016$ separable localized initial conditions (qubit at the origin)
and we also keep track of the number of initial conditions leading to highly entangled states after $1000$ steps.
The results of these simulations can be seen in Figs. \ref{fig2} and \ref{fig3}, where we have
the same features highlighted in Fig.~\ref{fig1}.

Now, however, the asymptotic limit is more clearly depicted and we undoubtedly see that
$\lim_{t\rightarrow \infty}\langle S_E(t)\rangle = 1$ for dynamic and fluctuating disorder and that 
static disorder has no asymptotic limit, with $\langle S_E(t)\rangle$ oscillating about a lower value.
For the ordered case, we now perceive that the oscillations have smaller amplitudes as the time increases.
This result is in agreement with the fact that the ordered case has an asymptotic limit \cite{aba06,sal12,vie13}.

The insets in Figs.~\ref{fig2} and \ref{fig3} show that the great majority of initial conditions (more than $95\%$) give
$S_E(t)\geq 0.95$ for dynamic and fluctuating disorder while for order and static disorder less than $20\%$ achieve such a feat.
Also, the rate of states leading to $S_E(t)\geq 0.99$ for static disorder and order is extremely small.
For dynamic disorder, however, we have for the two classes of $RQRW$ at least a $50\%$ rate of getting at step $1000$ $S_E(t)\geq 0.99$.

\begin{figure}[!ht]
\includegraphics[width=8cm]{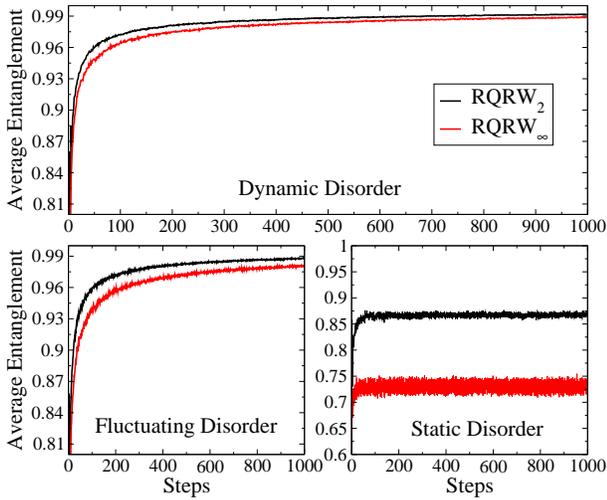}
\caption{\label{fig4}(color online) The same as Figs. \ref{fig2} and \ref{fig3} but now we want
to compare the entanglement generation efficiencies between $RQRW_\infty$ and $RQRW_2$. 
For all panels the upper/black curve represents $RQRW_2$ and the lower/red one $RQRW_\infty$, which
shows that $RQRW_2$ is more efficient than $RQRW_\infty$ for all types of disorder. This distinction
is more explicit for static disorder (lower-right panel).}
\end{figure}

Finally, in Fig. \ref{fig4} we plot the simulated data shown in Figs. \ref{fig2} and \ref{fig3} in such a way that we
are able to compare the entangling power between the two classes of $RQRW$ for a fixed type of disorder. 
As we can see in Fig. \ref{fig4}, for all types of disorder $RQRW_2$ outperforms $RQRW_\infty$.

\subsection{Special cases}

We now study how the entangling power of $RQRW_\infty$ and $RQRW_2$ is affected by reducing the disorder present in the system. 
Let us start by investigating how the entangling power of $RQRW_\infty$ changes if instead of having all three parameters $q$, $\theta$, and
$\varphi$ random, we fix two of them and let the remaining one be randomly chosen from a continuous uniform distribution. We also
compare how the entangling power is affected by successively decreasing the span of the distribution (less disorder), 
until we get very narrow ones, with the changing parameter barely deviating from the mean of the distribution (small variance). 
\begin{figure}[!ht]
\includegraphics[width=8cm]{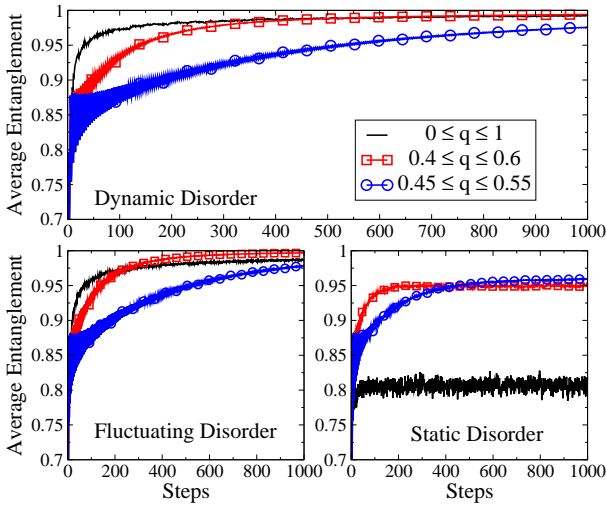}
\caption{\label{fig5}(color online) The curves were obtained averaging 
over 500 random initial conditions of the form,
$|\Psi(0)\rangle\hspace{-.1cm}=\hspace{-.1cm}(\cos\alpha_s|\hspace{-.1cm}\uparrow\rangle+e^{i\beta_s}
\hspace{-.1cm}\sin\alpha_s|\hspace{-.1cm}\downarrow\rangle) \otimes 
(\cos\alpha_p|-1\rangle+e^{i\beta_p}\hspace{-.1cm}\sin\alpha_p|1\rangle)$, 
where $\alpha_{s,p}\in [0,\pi]$ and $\beta_{s,p}\in [0,2\pi]$.  We worked with a $1000$-step walk
and with $RQRW_{\infty}$ where $\theta(j,t)=\varphi(j,t)=0$. 
Upper panel: Dynamic disorder with $q(j,t)=q(t)$ picked randomly at each time step 
from the three different uniform distributions listed in the graph. Lower-left panel: 
Fluctuating disorder with $q(j,t)$ picked randomly at each position and time from the same 
uniform distributions listed in the upper panel. Lower-right panel: Static disorder with
$q(j,t)=q(j)$ chosen at each position from the uniform distributions in the upper panel.}
\end{figure}
\begin{figure}[!ht]
\includegraphics[width=8cm]{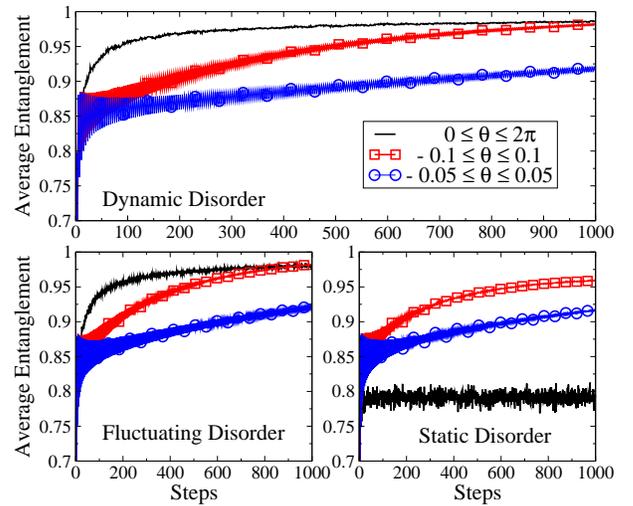}
\caption{\label{fig6}(color online) Same as explained in Fig.~\ref{fig5} but now 
$RQRW_{\infty}$ is such that $q(j,t)=1/2$, $\varphi(j,t)=0$, and $\theta(j,t)$ is the random
variable.}
\end{figure}
\begin{figure}[!ht]
\includegraphics[width=8cm]{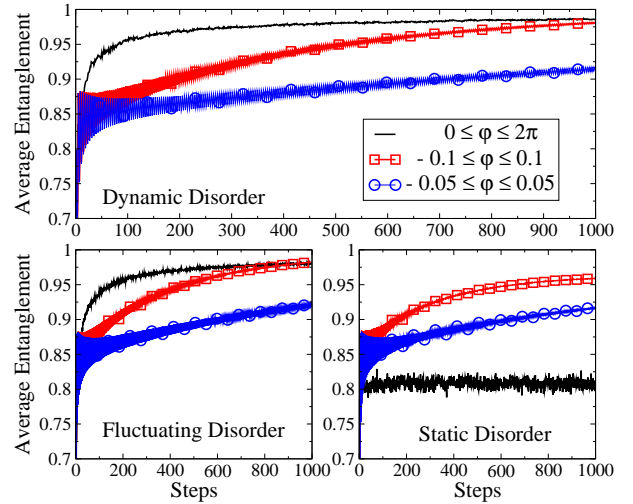}
\caption{\label{fig7}(color online) Same as in Fig.~\ref{fig5} but now 
$RQRW_{\infty}$ is such that $q(j,t)=1/2$, $\theta(j,t)=0$, and $\varphi(j,t)$
is the random parameter. }
\end{figure}

Looking at Figs. \ref{fig5}, \ref{fig6}, and \ref{fig7} we conclude that quantum walks with 
dynamic or fluctuating disorder are still
the most efficient entanglement generators and that as we reduce the span of the distribution from which the 
random variable is drawn, the rate at which the asymptotic limit is reached slows down. 
However, even those distributions with small variances furnish 
highly entangled states at the end of the walk, independently of the initial condition. For static disorder, a
surprising result appears when we reduce the span of the distribution. Indeed, in this case 
less disorder means greater entangling power and apparently the existence of a well defined asymptotic limit.
It is worth noticing that the interplay between the amount of disorder and the entangling power for the static case
is rather subtle. As the lower-right panels of Figs. \ref{fig5}, \ref{fig6}, and \ref{fig7} suggest, 
it is not clear how much disorder we must add to or reduce from the system in order to 
change its entangling power. There is probably a critical value for the amount of
disorder that we must consider in analyzing the static case. 
As we decrease disorder we start increasing the entangling power of 
the static $RQRW_\infty$. However, if we continue to decrease disorder beyond a critical value 
the system's entangling power starts to deteriorate. 
Further investigations that lie beyond the scope of this paper is needed to 
settle this matter satisfactorily.

\begin{figure}[!ht]
\includegraphics[width=8cm]{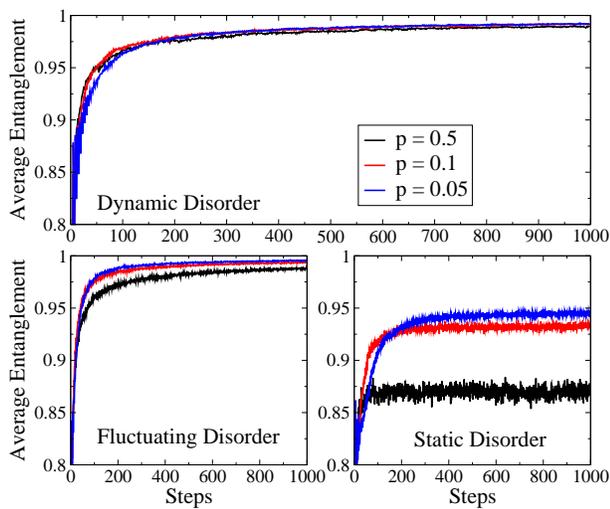}
\caption{\label{fig8}(color online) Same as in Fig.~\ref{fig5} but now 
we implement the biased $RQRW_2$, with $p$ being the probability of picking
the Hadamard coin and $(1-p)$ the Fourier coin. 
Upper panel: Dynamic disorder where at each time step $t$ the probability $p(j,t)=p(t)$ to pick 
the same Hadamard coin for all positions $j$ is given by the three values listed in the graph.
All curves tend to overlap, with the $p=0.1$ and $p=0.05$ giving slightly higher entanglement
for large $t$.
Lower-left panel: Fluctuating disorder where for each time $t$ and position $j$ the probability
$p(j,t)$ to get a Hadamard coin is given as depicted in the upper panel.
Here and in the next panel we have from top to bottom the $p=0.05$, $p=0.1$, and $p=0.5$ cases. 
Lower-right panel: Static disorder where the probability $p(j,t)=p(j)$ to get 
the Hadamard coin at position $j$ is as described in the upper panel. 
Note that for both the dynamic and fluctuating case, weak 
disorder is enough to guarantee $S_E\rightarrow 1$ asymptotically.}
\end{figure}

For $RQRW_2$ we studied how the bias between choosing the Hadamard or the Fourier/Kempe coin influences its entangling power.
In Fig. \ref{fig8} we see that again dynamic and fluctuating disorder lead to the 
most powerful entanglement generators, with their power being almost unaffected by changing the bias of choosing 
between the $H$ or $F$ coin. Indeed, for very low disorder, with the odds of drawing an $H$ coin at about
$5\%$, we still do not see appreciable changes in the entangling power for those walks 
when compared to the highly disordered (balanced) case. In the static case, we see a behavior similar to that
observed for $RQRW_\infty$. By decreasing disorder we have improved the entangling capacity of the walker considerably.

\subsection{Gaussian initial conditions}

\begin{figure}[!ht]
\includegraphics[width=8cm]{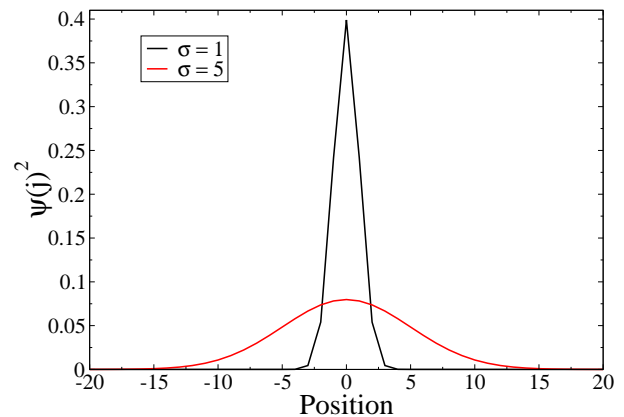}
\caption{\label{gauss}(color online) Discretized Gaussian distributions for the qubit's position.
The wider the Gaussian the greater $\sigma$.}
\end{figure}

We want now to study the entangling power of $RQRW_\infty$ for highly delocalized (non-local)
initial conditions in position. The initial positions we deal with are given by
Gaussian distributions centered about the origin, 
$$
\psi^2(x)=\frac{1}{\sqrt{2\pi\sigma^2}}e^\frac{-x^2}{2\sigma^2},
$$
with $\sigma^2$ being the variance. We set the total initial state (spin plus position ) as
$$
|\Psi(0)\rangle=|\xi\rangle\otimes \sum_{j=-\infty}^{\infty}\psi(j)|j\rangle,	
$$
with $\psi(j)\geq 0$ and $\psi(j)$ the discretized and normalized version of the 
Gaussian distribution, i.e., $\sum_{j}|\psi(j)|^2=1$.
We also employ two types of initial spin states,
$|\xi_1\rangle=(|\uparrow\rangle$ $+i|\downarrow\rangle)/\sqrt{2}$ and 
$|\xi_2\rangle=(|\uparrow\rangle+|\downarrow\rangle)/\sqrt{2}$. The two Gaussians
we work with are plotted in Fig.~\ref{gauss}.

\begin{figure}[!ht]
\includegraphics[width=8cm]{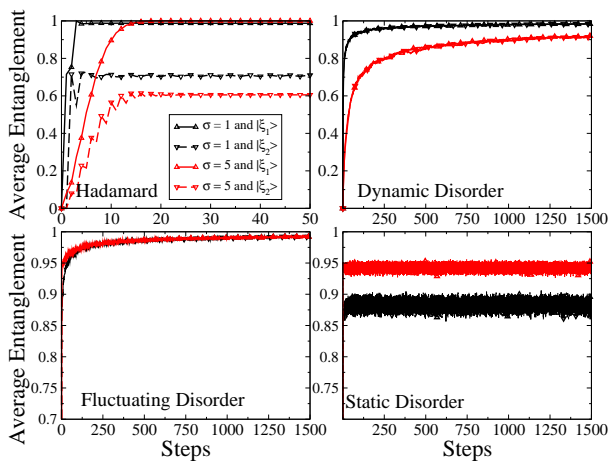}
\caption{\label{gaussSE}(color online) The solid curves give $\langle S_E(t) \rangle$ for 
Gaussian initial conditions with spin state $|\xi_1\rangle=(|\uparrow\rangle$ $+i|\downarrow\rangle)/\sqrt{2}$ 
and the dashed ones with 
$|\xi_2\rangle=(|\uparrow\rangle+|\downarrow\rangle)/\sqrt{2}$. The corresponding dispersions are given in the
upper-left panel, where we show $\langle S_E(t) \rangle$ for the Hadamard QRW (no averaging needed in this case).
Note that in this case the asymptotic entanglement is highly sensitive to the initial conditions.
In the other panels we have $RQRW_{\infty}$ with $q\in [0,1]$ and 
$\theta, \varphi \in [0,2\pi]$.}
\end{figure}

As we see in Fig. \ref{gaussSE} the ordered $QRW$ asymptotic entanglement is very sensitive to
the initial conditions, some of which not leading to a maximally entangled state. Static disorder is
also sensitive to the initial conditions of the qubit and neither furnishes maximally
entangled states nor has an asymptotic limit. 
Dynamic and fluctuating disorder are the most powerful entanglement generators with asymptotic behavior independent
of the initial conditions. For not too
delocalized initial states ($\sigma = 1$) these two types of disorder approach the asymptotic limit at the same rate. 
However, for highly delocalized states ($\sigma = 5$), fluctuating disorder has a faster rate. This is the only instance in 
which fluctuating disorder clearly outperforms dynamic disorder. Further investigations are needed, though, to definitively 
answer if for highly delocalized initial conditions fluctuating disorder is the optimal choice if we seek the fastest 
entangling rate.  

\subsection{The asymptotic limit}

The asymptotic limit is defined over the reduced density matrix $\rho_C(t)\in \mathcal{H}_C$, 
where $\rho_C(t)=Tr_P[\rho(t)]$ and $\rho(t) \in \mathcal{H}_C\otimes \mathcal{H}_P$.
In other words, it is defined over the spin space. Whenever a quantum walk leads to 
$$
\lim_{t\rightarrow \infty}D[\rho_C(t+1),\rho_C(t)]=0,
$$
we say the system has an asymptotic limit. Here $D(A,B)$ is any distance measure between
two operators $A$ and $B$ well defined in quantum mechanics.

To quantitatively study the asymptotic limit of the several walks here presented we
work with the distance called trace distance \cite{nie2000},
\begin{equation}
D[\rho_C(t+1),\rho_C(t)]= D(t) = \frac{1}{2}\text{Tr}\left(\left|\rho_C(t) - \rho_C(t-1)\right|\right),
\label{td}
\end{equation}
where  $|A| = \sqrt{A^\dagger A}$.  For a qubit a straightforward calculation shows
that $D(t)$ is equal to the Ky Fan 1-norm (largest singular value of 
$\rho_C(t) - \rho_C(t-1))$ and that the Frobenius norm of $\rho_C(t) - \rho_C(t-1)$ 
equals $\sqrt{2}D(t)$. Therefore, the results
we show below are essentially independent of what sort of distance we use. 

Computing explicitly $D(t)$ we get
\begin{equation}
D(t) = \frac{1}{2}\sqrt{[\Delta r_1(t)]^2+[\Delta r_2(t)]^2+[\Delta r_3(t)]^2},
\label{tdqubit}
\end{equation}
where $\Delta r_j(t) = \text{Tr}[\rho_C(t)\sigma_j]-\text{Tr}[\rho_C(t-1)\sigma_j]$
and $\sigma_j$ are the standard Pauli matrices.

It is known that the ordered Hadamard $QRW$ \cite{aba06} and the dynamic $RQRW$ \cite{vie13} 
have an asymptotic limit. Also, it is known that $D(t)$ goes to zero for those two cases 
obeying a power law with  different exponents \cite{vie13}. For the ordered case we
have $D(t)\sim t^{-1/2}$ while for dynamic disorder $D(t)\sim t^{-1/4}$.

Our goal in this section is to extend the analysis presented in \cite{vie13} 
and its supplemental material 
to fluctuating and static disorder. We want to check whether or not they have
an asymptotic limit and, in case of a positive answer, to compute the
rate at which $D(t)$ approaches zero.

\begin{figure}[!ht]
\includegraphics[width=8cm]{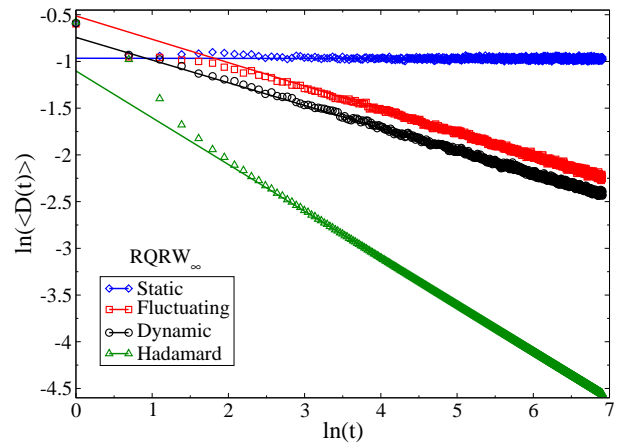}
\caption{\label{fig11}(color online)
The curves were obtained averaging 
over 1000 random initial conditions given as 
$|\Psi(0)\rangle\hspace{-.1cm}=\hspace{-.1cm}(\cos\alpha_s|\hspace{-.1cm}\uparrow\rangle+e^{i\beta_s}
\hspace{-.1cm}\sin\alpha_s|\hspace{-.1cm}\downarrow\rangle) \otimes 
(\cos\alpha_p|-1\rangle+e^{i\beta_p}\hspace{-.1cm}\sin\alpha_p|1\rangle)$, 
where $\alpha_{s,p}\in [0,\pi]$ and $\beta_{s,p}\in [0,2\pi]$.  
We work with $RQRW_{\infty}$ with $q$, $\theta$, and
$\varphi$ random and we go up to $1000$ steps.
The solid curves were obtained by a $95\%$ confidence
level linear fitting of the simulated data, where we have dropped the first $100$ points
to better highlight the asymptotic behavior. For dynamic and fluctuating disorder the
angular coefficients are respectively ($-0.2429 \pm 0.0009$) and ($-0.2500 \pm 0.0008$) while for the ordered 
Hadamard $QRW$ it is ($0.5005 \pm 0.0001$). The static case
gives a parallel line to the $x$ axis with angular coefficient ($-0.0006\pm 0.0006$).
These results corroborate the fact that whenever we have randomness in time (dynamic and fluctuating cases)
the distance between successive matrices $\rho_C(t)$ tend to zero as $D(t)\sim t^{-1/4}$ while for
the ordered case $D(t)\sim t^{-1/2}$. For static disorder $\rho_C(t)$
does not approach a particular state and we do not have an asymptotic limit ($\lim_{t\rightarrow\infty}D(t)\neq 0$
or, equivalently, $\lim_{t\rightarrow\infty}\ln[D(t)]\nrightarrow -\infty$).
}
\end{figure}

In order to determine precisely whether or not we have an asymptotic limit and the 
rate at which $D(t)\rightarrow 0$ as we increase $t$, we calculated $D(t)$ for hundreds of 
random initial conditions and computed its average $\langle D(t)\rangle$. We observed that
for all cases where an asymptotic limit exists, $\langle D(t)\rangle = at^b$. Hence,
since $\ln [\langle D(t)\rangle] = \ln(a) +b\ln(t)$, where $\ln$ is the natural
logarithm, the angular coefficient of the line in a 
$\ln [\langle D(t)\rangle]$ versus $\ln(t)$ plot gives the coefficient $b$ characterizing the
power law associated with $D(t)$.

\begin{figure}[!ht]
\includegraphics[width=8cm]{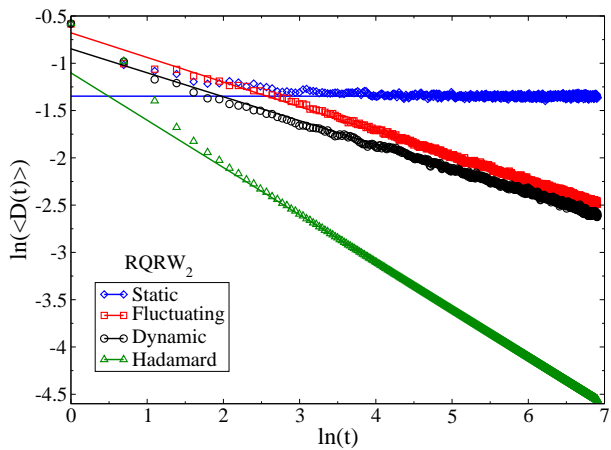}
\caption{\label{fig12}(color online)
Same as Fig. \ref{fig11} but now we have the balanced $RQRW_2$, 
with Fourier and Hadamard coins.
For dynamic and fluctuating disorder the
angular coefficients are respectively ($-0.2533 \pm 0.0009$) and ($-0.2604 \pm 0.0008$). 
The static case gives a parallel line to the $x$ axis with angular coefficient ($-0.0002\pm 0.0007$).
The simulated data once again corroborate the fact that whenever we have randomness in time (dynamic and fluctuating cases)
the distance between successive matrices $\rho_C(t)$ tend to zero as $D(t)\sim t^{-1/4}$ while for
the ordered case $D(t)\sim t^{-1/2}$.}
\end{figure}

\begin{figure}[!ht]
\includegraphics[width=8cm]{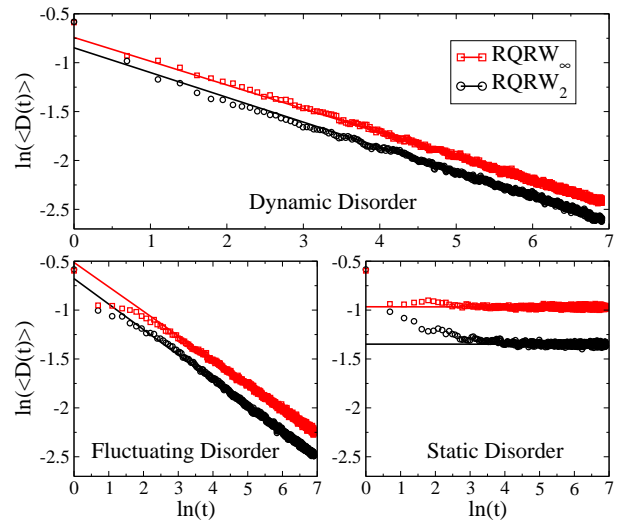}
\caption{\label{fig13}(color online)
The same configuration as given in Figs. \ref{fig11} and \ref{fig12}. This time we compare
within a given type of disorder the behavior of $\langle D(t)\rangle$ for 
$RQRW_{\infty}$ and $RQRW_2$.}
\end{figure}

In Fig. \ref{fig11} we show the results of this analysis for $RQRW_\infty$ and in Fig.~\ref{fig12} 
for $RQRW_2$. As we can see from the simulated data, fluctuating disorder has a clear asymptotic
limit for both types of walks, $RQRW_\infty$ and $RQRW_2$, while static disorder has no asymptotic
limit. Also, we have determined that $D(t)\sim t^{-1/4}$ for fluctuating disorder, the same power law
when we have dynamic disorder. For completeness, we also show the curves for dynamic disorder 
($D(t)\sim t^{-1/4}$) and for the ordered case ($D(t)\sim t^{-1/2}$), 
which corroborate the conclusions given in \cite{vie13}. In Fig. \ref{fig13} we compare the
time behavior of $D(t)$ between $RQRW_\infty$ and $RQRW_2$ within a given type of disorder.
We can see that $RQRW_2$ has slightly lower linear coefficients for all types of disorder.

\subsection{Transport properties}

The transport properties of ordered and some types of disordered quantum walks were already studied theoretically and experimentally
in many previous works \cite{aha93,kem03,and12,car05,aba06,sal12,bos09,goy10,fra11,shi10,shi13,shi14,all12,rol13,mou13,rib04,ban06,bue04,woj04,rom05,sha03,joy11,
ahl11a,ahl11b,ahl12,vie13, sch10,sch11,cre13,cha12} and here we just highlight the main results concerning the particular types 
of $RQRW$ we have been working with.
\begin{figure}[!ht]
\includegraphics[width=8cm]{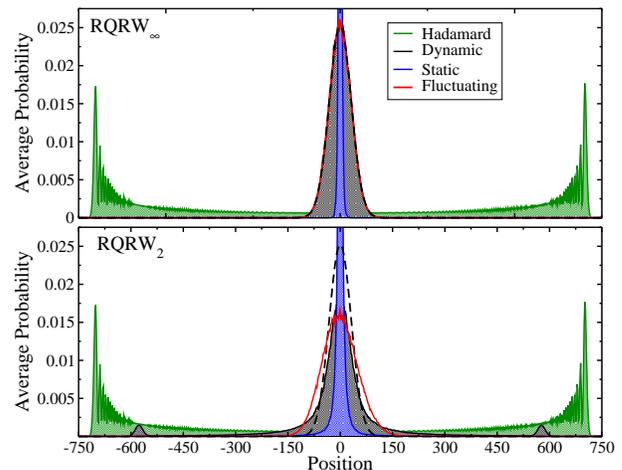}
\caption{\label{fig14}(color online)
Average probability distributions $\langle P(j)\rangle$ for all $RQRW$ described in
Figs. \ref{fig11} (upper panel) and \ref{fig12} (lower panel) after $1000$ steps. The black/dashed curves
represent the expected probability distribution for the balanced $CRW$ starting at the origin and subjected
to the same number of steps of the quantum processes. Upper panel: Dynamic (black/solid) and
fluctuating (red/solid) curves overlap with the classical one (black/dashed). 
Lower panel: For $RQRW_2$, dynamic disorder still has spikes away from the origin, while this feature
is no longer present for fluctuating disorder. In both panels static disorder (blue/solid) gives
distributions that are narrower than the classical one (Anderson localization).}
\end{figure}

In Fig. \ref{fig14} we plot the average probability distribution $\langle P(j) \rangle$
over $1000$ realizations, with each realization starting  at a different initial condition. 
The quantity $\langle P(j) \rangle$ 
gives the average probability to find the qubit at position $j$ after a predefined number of steps. In Fig. \ref{fig14} we
show $\langle P(j) \rangle$ after $1000$ steps. 
 
\begin{figure}[!ht]
\includegraphics[width=8cm]{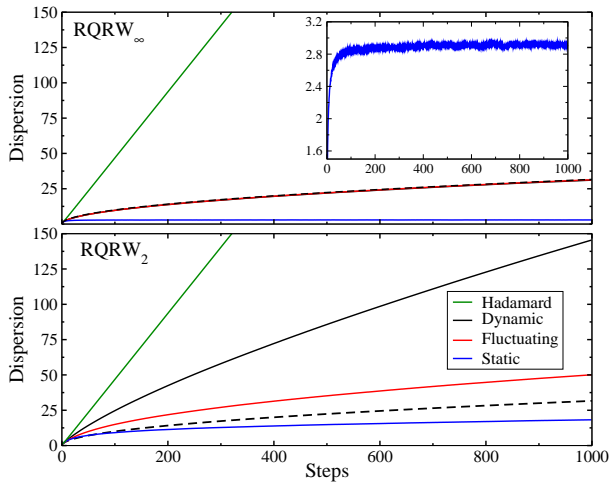}
\caption{\label{fig15}(color online)
Average dispersions $\langle \sqrt{\sigma^2}\rangle$ for all $RQRW$ described in
Figs. \ref{fig11} (upper panel) and \ref{fig12} (lower panel). The black/dashed curves
represent the expected dispersion for the balanced $CRW$. Upper panel: Dynamic (black/solid) and
fluctuating (red/solid) curves overlap with the classical dispersion (black/dashed)
and are given as $\sigma \sim \sqrt{t}$ (diffusive behavior). 
The Hadamard $QRW$ curve (green/solid) is such that $\sigma \sim t$ 
(ballistic behavior). The static case (blue/solid) has $\sigma \sim constant$ (see inset) 
and is far below all curves, clearly indicating Anderson localization. 
Lower panel: For $RQRW_2$, all curves are distinct with
dynamic and fluctuating disorder lying between the ordered and the classical case. 
Static disorder dispersion is no longer a constant but still lies below all curves.}
\end{figure}

In the upper panel of Fig. \ref{fig14} we work with $RQRW_\infty$. Looking at the curves for $\langle P(j) \rangle$,
we note that the dynamic and fluctuating disorder curves are barely
indistinguishable from the expected probability distribution for the unbiased classical walk starting at the origin, indicating
a diffusive behavior for these two types of disorder. This is confirmed by looking at the upper panel of Fig. \ref{fig15},
where the average dispersion for dynamic and fluctuating disorder behaves like $\sigma \sim \sqrt{n}$. The dispersion
curves for these two types of disorder are also indistinguishable from the classical dispersion. 
For static disorder, $\langle P(j)\rangle$ is narrower than the classical curve, which indicates Anderson localization. 
This is proved true by looking at the dispersion curve in the upper panel of Fig. \ref{fig15}, and in particular in the inset,
where it is clear that its long time behavior is $\sigma \sim constant$, the key signature of Anderson localization.

The analysis for $RQRW_2$ can be seen at the lower panels of Figs.~\ref{fig14} and \ref{fig15}.
The overall transport properties of $RQRW_2$ are similar to those of $RQRW_\infty$. The main difference is related
to the fact that the probability distribution curves and the dispersion curves are all distinguishable from
each other. Moreover, dynamic and fluctuating disorder have no longer a diffusive behavior, being more delocalized 
than the classical case. 
Indeed, for dynamic disorder we still see non-negligible probabilities of seeing the
qubit away from the origin (see the spikes around site $j=\pm 550$ at the lower panel of Fig.~\ref{fig14}). 
Note, however, that they are still by far more localized than the ordered case.
Also, static disorder probability distribution curve is still narrower than
the classical one, but does not achieve perfect Anderson localization after $1000$ steps. This can be seen by
looking at the dispersion curve. Although it is below the classical one, is not yet a constant 
(see lower panel of Fig. \ref{fig15}).

\section{Discussions and Conclusion}

In this article we have addressed how different types of disorder affect the entangling power of the simplest quantum
version of the one-dimensional classical random walk ($CRW$), namely, the discrete time quantum random walk ($QRW$), 
which is built on a single qubit that moves right or left according to its internal degree of freedom (spin). 
During its walk, the internal degree of freedom of the system gets entangled with the external one (position).
For the ordered case, this entanglement is highly sensitive to the initial conditions and almost never reaches the
maximal value possible.

To introduce disorder in $QRW$ we allow the quantum coin (an $SU(2)$ unitary matrix) to be randomly chosen at each
time step and/or lattice position. Following \cite{vie13}, this extra random aspect (disorder) led us to call such 
processes random quantum random walks ($RQRW$).
The three types of disorder we investigated are called dynamic, fluctuating, and static disorder and they appear due to
imperfections in the construction of the quantum coin and to inherently quantum fluctuations during the time evolution of the qubit.
Which one dominates a given $RQRW$ depends on the interplay between the rate of the fluctuations in the system and the total time of 
the experiment. They were already implemented in the laboratory and the level and type of disorder present in the system can also
be tuned to almost any desired value \cite{sch10,sch11,cre13}. 

In dynamic $RQRW$ the quantum coins are the same throughout the lattice, changing randomly at each time step. 
In the static $RQRW$ the quantum coins are fixed in time but are once randomly distributed along the lattice sites.
Fluctuating $RQRW$ combines both features of the previous cases, the quantum coins change randomly at each time step and at each
position. 

So far the entangling power of $RQRW$ is only extensively known for dynamic disorder \cite{vie13} where,
contrary to the ordered case, it leads asymptotically in the number of steps
to maximally entangled states for any initial condition. Little is known about the entangling power of static \cite{cha12}
and fluctuating disorder. In this paper we filled this gap by investigating in depth the entangling power of the 
last two aforementioned types of disorder.  

By means of extensive numerical studies for several thousands of initial conditions and many different levels of disorder, we showed that
dynamic disorder is almost always the strongest entangler, being followed closely by fluctuating disorder. Both types of disorder give
maximally entangled states asymptotically and independently of the initial conditions. Static disorder, however, leads to the worst results in terms of 
entanglement generation, being almost always worse than the ordered case. Only when a low level of static disorder is added to the
system does it outperform the ordered case, but still losing to the dynamic and fluctuating cases. 
Also, the static $RQRW$ has no asymptotic limit, with the long time value of entanglement fluctuating
about a mean value with no signs of convergence to its mean. Furthermore, the reduced density matrix 
describing the internal degree of freedom of the qubit
fluctuates randomly among several different matrices, while for dynamic \cite{vie13} and fluctuating disorder 
it tends to a specific one 
(to the asymptotic value $\lim_{t\rightarrow \infty}\rho_C(t)$).

We have also studied the rate at which the asymptotic limit is approached, i.e., the rate at which $\rho_C(t)$
approaches its asymptotic value. As already noted for dynamic disorder \cite{vie13}, we observed that 
for fluctuating disorder the asymptotic limit is approached according to a power law, $t^{-1/4}$, whose 
exponent is the same as the one for dynamic disorder. For the ordered case we also have a power law but with 
different exponent, $t^{-1/2}$.  

Putting together the results obtained here and in \cite{vie13} and its supplemental material, namely, (i) 
static disorder has no asymptotic limit and is the worst entanglement generator; 
(ii) dynamic and fluctuating disorder give asymptotically
maximally entangled states for any initial condition; and (iii) deterministic/periodic time-dependent coins \cite{vie13} do not 
generate highly entangled states for all initial conditions, we are led to conclude that we must have at least the following two ingredients to
produce quantum walks such that we get maximally entangled states for any initial condition: We must have time-dependent
coins changing randomly throughout the time evolution. Indeed, time-independence with position-dependent randomness (static disorder) or
time-dependence with no randomness do not lead to maximal entanglement for all initial conditions. 
For dynamic and fluctuating disorder, however, where their common feature
is time-dependent randomness, we do get maximally entangled states for any initial condition if we wait enough.

\begin{acknowledgments}
RV thanks CAPES (Brazilian Agency for the Improvement of Personnel of Higher Education)
for funding and GR thanks the Brazilian agencies CNPq
(National Council for Scientific and Technological Development) and
FAPESP (State of S\~ao Paulo Research Foundation) for funding and
CNPq/FAPESP for financial support through the National Institute of
Science and Technology for Quantum Information.
\end{acknowledgments}

\end{document}